\documentclass[11pt,preprint]{aastex}

\def\etal{{et~al.}}

\def\lae{\mathrel{<\kern-1.0em\lower0.9ex\hbox{$\sim$}}}
\def\gae{\mathrel{>\kern-1.0em\lower0.9ex\hbox{$\sim$}}}

\def\etal{et~al.~}

\shorttitle{The Central Brightness Profiles of Early-Type Galaxies}

\shortauthors{C\^ot\'e \etal}

\begin{document}

\title{The ACS Fornax Cluster Survey. II. The Central Brightness Profiles of Early-Type Galaxies:
A Characteristic Radius on Nuclear Scales and the Transition from Central
Luminosity Deficit to Excess\altaffilmark{1}}

\author{Patrick C\^ot\'e\altaffilmark{2},
Laura Ferrarese\altaffilmark{2},
Andr\'es Jord\'an\altaffilmark{3},
John P. Blakeslee\altaffilmark{4},
Chin-Wei Chen\altaffilmark{2,5},
Leopoldo Infante\altaffilmark{6},
David Merritt\altaffilmark{7},
Simona Mei\altaffilmark{8},
Eric W. Peng\altaffilmark{2},
John L. Tonry\altaffilmark{9},
Andrew~A.~West\altaffilmark{10},
Michael~J.~West\altaffilmark{11}}

\altaffiltext{1}{Based on observations with the NASA/ESA {\it Hubble
Space Telescope} obtained at the Space Telescope Science Institute,
which is operated by the Association of Universities for Research in
Astronomy, Inc., under NASA contract NAS 5-26555.}
\altaffiltext{2}{Herzberg Institute of Astrophysics, National
Research Council of Canada, Victoria, BC, V9E 2E7, Canada}
\altaffiltext{3}{European Southern Observatory, Karl-Schwarzschild-Str.
2, 85748 Garching, Germany}
\altaffiltext{4}{Department of Physics \& Astronomy, Washington State University,
Pullman, WA 99164-2814}
\altaffiltext{5}{Institute of Astronomy, National Central University
Taiwan, 32054, Chungli, Taiwan}
\altaffiltext{6}{Departamento de Astronom\'{\i}a y Astrof\'{\i}sica, 
Pontificia Universidad Cat\'olica de Chile, Santiago 22, Chile} 
\altaffiltext{7}{Department of Physics, Rochester Institute of
Technology, 84 Lomb Memorial Drive, Rochester, NY 14623}
\altaffiltext{8}{ GEPI, Observatoire de Paris, 
Meudon Cedex, France}
\altaffiltext{9}{Institute for Astronomy, University of Hawaii, 2680
Woodlawn Drive, Honolulu, HI 96822}
\altaffiltext{10}{Department of Astronomy, University of California,
601 Campbell Hall, Berkeley, CA 94720}
\altaffiltext{11}{Department of Physics \& Astronomy, University of
Hawaii, Hilo, HI 96720; and Gemini Observatory, Casilla 603, La Serena, Chile}

\slugcomment{To appear in the {\it Astrophysical Journal}, December 2007}

\begin{abstract}
We analyse brightness profiles for 143 early-type galaxies in
the Virgo and Fornax Clusters, observed with the {\it Advanced Camera for Surveys}
on the {\it Hubble Space Telescope}. S\'{e}rsic models are found to
provide accurate representations of the global profiles with a notable exception: the observed
profiles deviate systematically inside a characteristic ``break" radius
of $R_b \approx 0.02^{+0.025}_{-0.01}R_e$,
where $R_e$ is the effective radius of the galaxy. The sense of the
deviation is such that bright galaxies ($M_B \lesssim -20$) typically show
central light {\it deficits} with respect to the inward extrapolation of the
S\'{e}rsic model, while the great majority of low- and intermediate-luminosity
galaxies ($-19.5 \lesssim M_B \lesssim -15$) show central light {\it excesses};
galaxies of intermediate luminosities
($-20 \lesssim M_B \lesssim -19.5$) are generally well fitted by S\'{e}rsic
models over all radii. We show that the slope, $\gamma^{\prime}$, of the central surface
brightness profiles, when measured at fixed fractions of $R_e$,
varies smoothly as a function of galaxy luminosity in a manner that depends
sensitively on the choice of measurement radius. We find no evidence for a
core/power-law dichotomy, and show that a recent claim of
strong bimodality in $\gamma^{\prime}$ is likely  an artifact of the biased galaxy selection
function used in that study.  To provide a more robust characterization of
the inner regions of galaxies, we introduce
a parameter, ${\Delta}_{0.02} = \log{({\cal L}_g/{\cal L}_s)}$ --- where
${\cal L}_g$ and ${\cal L}_s$ are the integrated luminosities inside
$0.02R_e$ of the observed profile and of the inward extrapolation of the outer
S\'ersic model --- to describe the central luminosity deficit (${\Delta}_{0.02} < 0 $) or excess
(${\Delta}_{0.02} > 0 $). We find that ${\Delta}_{0.02}$ varies smoothly over
the range of $\approx$ 720~in luminosity spanned by the
sample galaxies, with again no evidence for a dichotomy.
We argue that the central light excesses in $M_B \gtrsim -19$
galaxies may be the analogs of the dense central cores predicted
by some numerical simulations to form via gas inflows.
\end{abstract}

\keywords{galaxies: clusters: individual (Virgo, Fornax)--galaxies:
elliptical and lenticular, cD--galaxies: nuclei: galaxies: structure}

\section{Introduction}
\label{sec:intro}

Pioneering HST imaging studies of the centers of early-type galaxies suggested an apparently abrupt
transition in central stellar density at $M_B \sim -20.3$ mag --- the so-called ``core/power-law
dichotomy" (e.g., Ferrarese \etal 1994; Lauer \etal 1995). These findings prompted the widely
held view that the bright (``core") and faint (``power-law") galaxies follow distinct
evolutionary routes (e.g., Faber \etal 1997). However, the evidence for such a dichotomy has lessened
(although not entirely disappeared) following more recent studies that identified a 
population of galaxies with intermediate properties (Rest \etal 2001; Ravindranath \etal 2001).
The slope, $\gamma^{\prime}$, of the central surface brightness profile --- usually parameterized as a
``Nuker" law (essentially two power-laws that merge at a characteristic ``break" radius;
Lauer \etal 1995) and measured at the angular distance corresponding to the instrumental
resolution --- has traditionally been taken as a diagnostic of this behavior.

However, using ACS profiles for 100 early-type galaxies belonging to the Virgo Cluster, Ferrarese \etal (2006a)
showed that S\'{e}rsic models provide more accurate parameterizations of the global brightness
profiles than do Nuker models (see also Graham 2004; Ferrarese \etal 2006c) and 
argued that the core/power-law dichotomy is an artifact introduced in part by the
use of an inappropriate (i.e., power-law) parameterization of the outer profiles,
combined with a tendency in previous work (which usually relied on HST brightness 
profiles of limited radial extent) to not properly account for
the compact stellar nuclei found in low- and intermediate luminosity galaxies (e.g.,
Graham \& Guzm\'an 2003; Grant \etal 2005; C\^ot\'e \etal 2006). 

In this paper, we use the best available imaging dataset --- in terms of depth, radial
coverage, angular resolution, completeness and homogeneity --- to re-examine the
central structure of early-type galaxies. Our analysis relies on HST/ACS imaging for
100 early-type members of the Virgo (previously discussed in Ferrarese \etal 2006ab and
C\^ot\'e \etal 2006) and new HST/ACS imaging for 43 early-type members of the Fornax
cluster (Jord\'an \etal 2007).
Our principle finding is clear evidence for a continuous, systematic progression
from central luminosity deficit ($M_B \lesssim -20$) to excess ($M_B \gtrsim -19$) within a
characteristic radius, approximately equal to 2\% of the galaxy effective radius. We
find no evidence for a ``core/power-law" dichotomy.

\section{Observations}

{\it HST} images for 143 members of the Virgo and Fornax Clusters were
acquired with the Advanced Camera for Surveys (ACS, Ford \etal\ 1998)
as part of the ACS Virgo (ACSVCS; GO-9401) and Fornax (ACSFCS; GO-10217)
Cluster Surveys (C\^ot\'e \etal 2004; Jord\'an \etal 2007). Surface brightness
fluctuation distance measurements (Mei \etal 2005, 2007; Blakeslee \etal 2008, in preparation) reveal the program
galaxies to span luminosity ranges of $\approx$ 545 (Virgo), $\approx$ 345
(Fornax) and $\approx$ 720 (combined). All galaxies have early-type morphologies (i.e., E, S0, dE,
dE,N or dS0) and are confirmed velocity members of their respective clusters. 
Images were taken in the Wide Field Channel (WFC) mode with a filter combination
(F475W and F850LP) roughly equivalent to the $g$ and $z$ bands in the
Sloan Digital Sky Survey (SDSS) photometric system.  The images cover a roughly
200$^{\prime\prime}\times200^{\prime\prime}$ field with 0\farcs05 pixel$^{-1}$ sampling
and $\approx 0\farcs1$ resolution. This resolution limit translates
to physical scales of 8.0 and 9.5~pc for Virgo and Fornax, respectively.
For the 21 Virgo galaxies brighter than $B_T = 12~(M_B \approx -19.2)$, the profiles were 
extended in radius by matching the ACS profiles to those measured
from SDSS $g$ and $z$ mosaics (Data Release 5; Adelman-McCarthy \etal 2007)
generated using the procedures described in West \etal (2007). The ACSVCS
sample is complete for early-type galaxies brighter than $B \approx 12$
($M_B \approx -19.2$) and 44\% complete down to its limiting magnitude of
$B \approx 16$ ($M_B \approx -15.2$). The ACSFCS sample, meanwhile, is
complete down to its limiting magnitude of $B \approx 15.5$
($M_B \approx -16.1$) with the exception of a single galaxy (FCC161 = NGC1379)
for which the data acquisition failed because of a shutter problem.

Full details on the construction of the azimuthally averaged brightness profiles and the adopted
fitting procedures (e.g., correction for dust obscuration, masking of background
sources, the identification of offset nuclei via centroid shifts, and the choice of
weighting schemes and minimization routines, etc) are given in \S\S 3.1, 3.2 and 3.3 
of Ferrarese \etal (2006a) and \S\S 3.1, 3.2, 4.1 and 4.2 of C\^ot\'e \etal\ (2006).

\subsection{Parameterization of the Brightness Profiles}
\label{sec:param}

Figures~1 and 2 shows $g$ and $z$ surface brightness profiles for nine representative
galaxies belonging to each of the ACSVCS and ACSFCS 
samples. To parameterize the brightness profiles, we begin by noting
that S\'{e}rsic (1968) models
\begin{equation}
I_S(R) = I_e \exp \left\{ -b_n \biggl [ \biggl ({R \over R_e} \biggl)^{1/n} - 1 \biggr ] \right\},
\end{equation}
generally provide accurate descriptions of the global profiles of the program
galaxies --- most importantly the downward curvature on large scales --- with
only three free parameters ($I_e$, $n$ and $R_e$). However, the
{\it central} regions of the profiles (typically $R \lesssim 100$ pc for the brightest
galaxies, and $R \lesssim 10$ pc for the faintest), deviate significantly from the inward
extrapolations of the S\'{e}rsic profiles for both galaxies brighter
than $M_B \lesssim -20$ (which show central light {\it deficits}) and for 
most galaxies fainter than $M_B \gtrsim -19.5$ (which usually show
central {\it excesses}). By contrast, most galaxies with $-20 \lesssim M_B \lesssim -19.5$
are reasonably well fitted by a single S\'{e}rsic model over all
radii, as previously noted by Graham \& Guzm{\'a}n (2003). The discovery of
a systematic progression from central light deficit to excess along the
luminosity function was discussed extensively in C\^ot\'e \etal (2006) and
Ferrarese \etal\ (2006ab). This transition from deficit to excess at $-20.0 \lesssim M_B \lesssim -19.5$ corresponds
to S\'ersic indices of $3.9 \lesssim n \lesssim 3.5$, based on the scaling relations presented in \S4.3 of 
Ferrarese \etal\ (2006a; see their Eq. 27).

\begin{figure}
\plotone{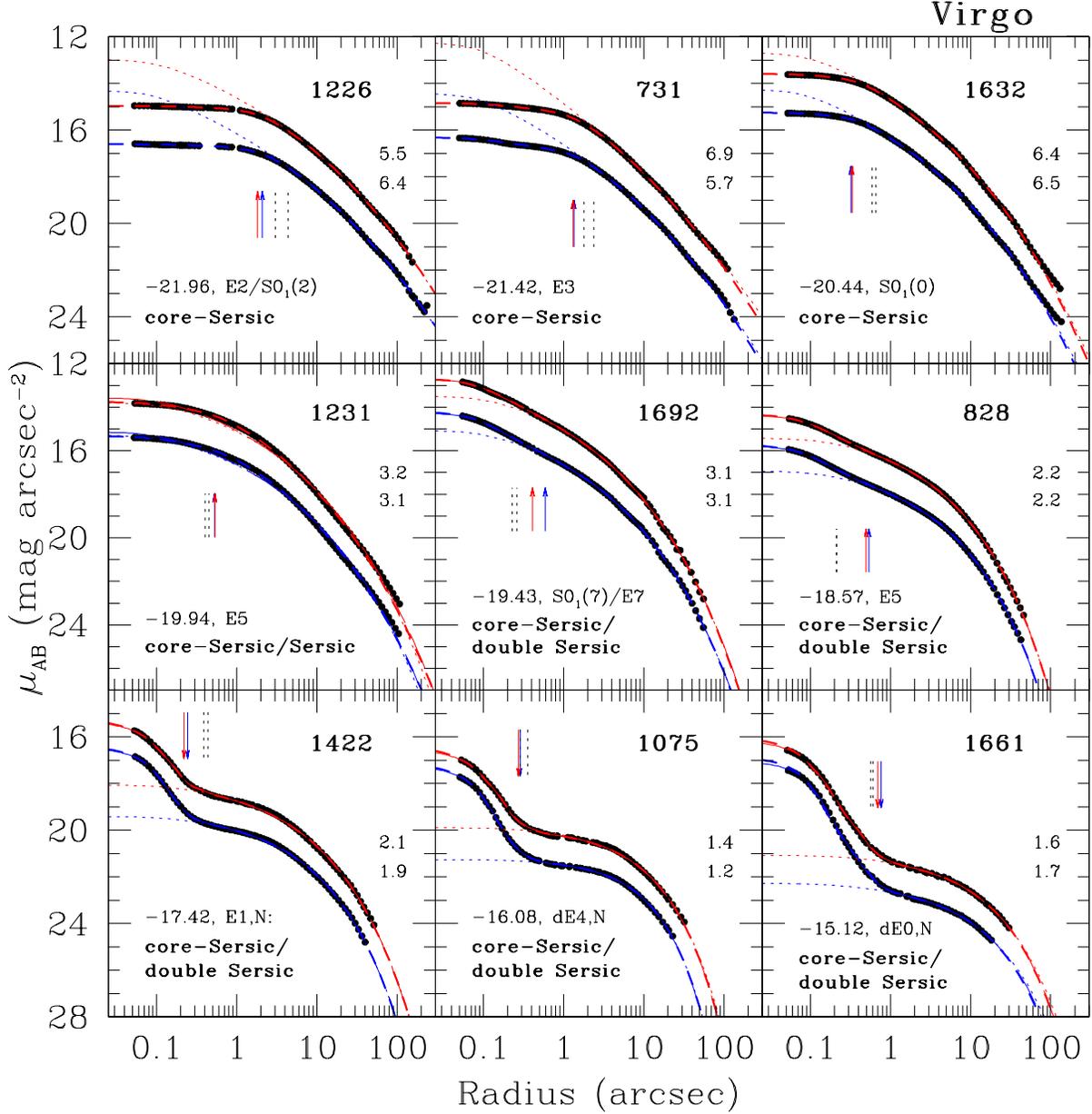}
\figurenum{1}
\caption{Surface brightness profiles for nine representative
galaxies from the ACSVCS. Identification numbers from the Virgo Cluster 
Catalog (VCC) of Binggeli, Sandage \& Tammann (1985) are given in the upper
right corner of each panel. For each of these galaxies, which span a 
range of $\approx$ 545~in blue luminosity, we show both the $g$ and $z$
profiles as the lower and upper points, respectively. The galaxies are ordered
according to decreasing absolute blue magnitude which is recorded in each panel,
along with their morphological types and the best-fit S\'ersic indices, $n$, for the
galaxy measured in the $g$ and $z$ profiles (lower and upper labels). For each profile, 
we show the best-fit ``core-S\'{e}rsic" (short dashed curves) and ``composite" model
(long dashed curves; see \S\ref{sec:param}). The red and blue
arrows show the break radii, $R_b$, for the fitted core-S\'{e}rsic models (in $g$ and
$z$, respectively) while the dotted vertical lines in each panel are drawn at 2\%
of the effective radius.
The dotted curves show the inward extrapolations of the
S\'{e}rsic component that best fits the profile for $R \gtrsim R_b$. Note the smooth
transition from luminosity ``deficit" to ``excess" as one moves
down the luminosity function.
\label{fig:sbp1}}
\end{figure}

\begin{figure}
\plotone{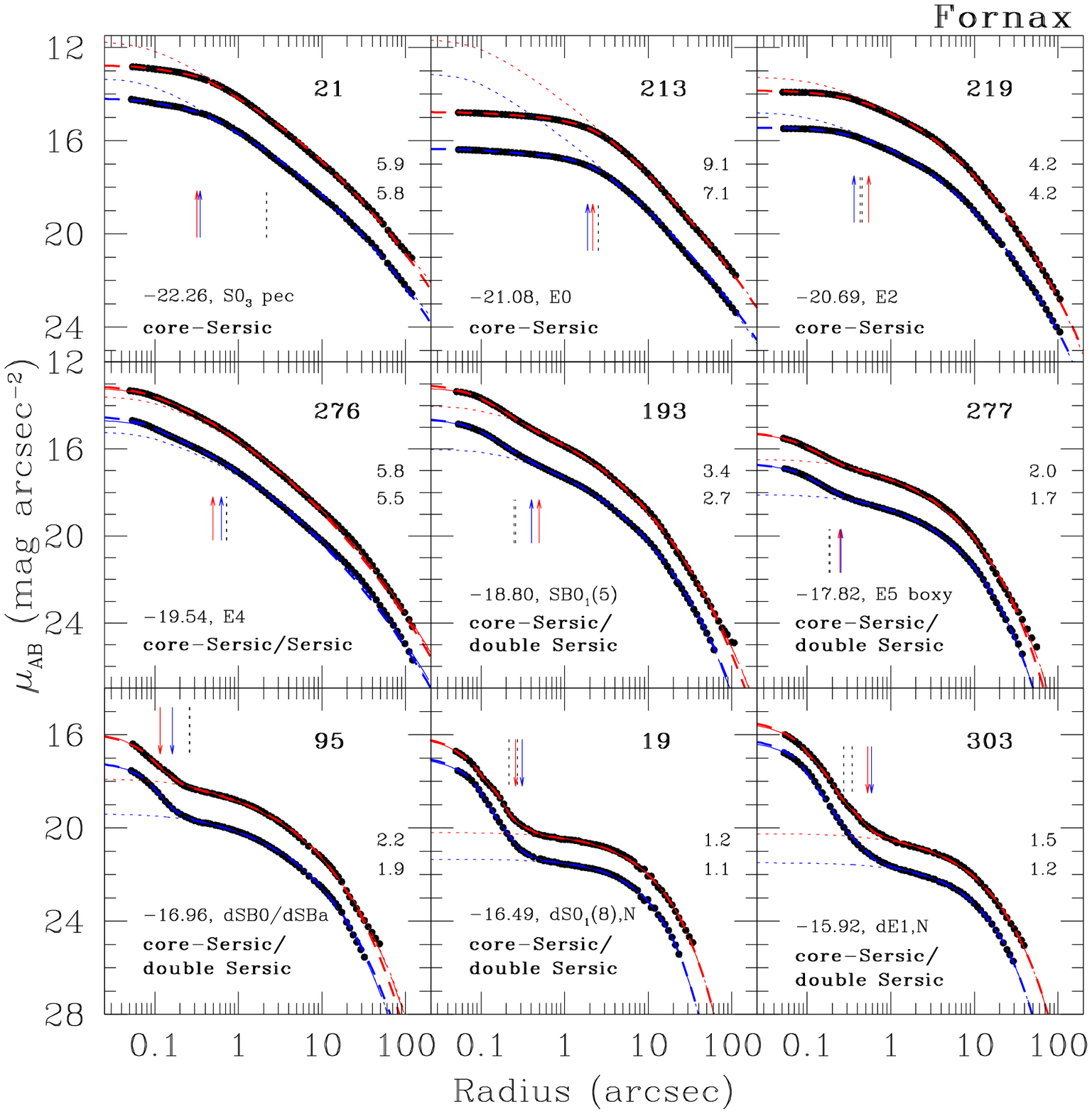}
\figurenum{2}
\caption{Same as in the previous figure, except for nine representative galaxies from the
ACSFCS. Identification numbers from the Fornax Cluster Catalog (FCC) of Ferguson (1989).
The galaxies, which span a range of $\approx$ 345~in blue luminosity, exhibit the same
basic trends as those in the Virgo Cluster. Note the small break radius for 
FCC21 (= NGC1316 = Fornax A), a peculiar S0 LINER/radio galaxy that shows clear evidence
of recent merging (i.e., dust, H$\alpha$ filaments, shells and ripples; Schweizer 1980).
\label{fig:sbp2}}
\end{figure}

To put this trend on a more quantitative footing, we parameterize the
observed brightness profiles of the ACSVCS and ACSFCS galaxies in two different ways.
First, given that the departures from S\'{e}rsic models always 
occur on scales of a few arcseconds and usually much less --- meaning that the 
innermost behavior is dominated by the instrumental point spread function (PSF) ---
a simple, PSF-convolved, power-law profile (with
$0 \lesssim \gamma^{\prime} \lesssim 2$) inside a break radius, $R_b$, 
usually provides an adequate representation of the observed profiles. The
short dashed curves in each panel of Figures~1 and 2 shows such ``{\tt core-S\'{e}rsic}"
models (Graham \etal 2003),
\begin{equation}
I_{cS}(R) = I^{\prime} \biggl [1 + \biggl ( {R_b \over R} \biggr )
^{\alpha} \biggl ]^{\gamma / \alpha} \exp \biggl [-b_n  \biggl (
{R^{\alpha} + R_b^{\alpha} \over R_e^{\alpha}} \biggr ) ^{1/(\alpha
n)} \biggr ],
\end{equation}
fitted to the profiles in both bandpasses after convolution with the instrumental PSFs. 
Note that $I^{\prime}$ in Eq.~(2) is related to the intensity, $I_b$,  at the break
radius $R_b$ through the relation:
\begin{equation}
I^{\prime} = I_b2^{-\gamma / \alpha} \exp \biggl [b_n \biggl
(2^{1/\alpha}R_b/R_e \biggl )^{1/n} \biggr ].
\end{equation}
Thus, this choice of parameterization requires a total of five free
parameters ($I_b$, $\alpha$, $\gamma$, $n$ and $R_e$).  For comparison, the
dotted curves in each panel of Figures~1 and 2 show the inward extrapolations of an outer S\'{e}rsic
component, illustrating the systematic evolution from deficit to
excess. The arrows show the fitted values
of $R_b$ for each galaxy.

As pointed out above, the profiles in the innermost regions are
dominated by the instrumental PSF, so a different parameterization
of the inner component could provide equally acceptable fits,
particuarly for the low- and intermediate-luminosity galaxies. Indeed, a detailed study
of the nuclear brightness profiles of nearby galaxies (Ferrarese \etal 2008, in
preparation) suggests that an excellent parameterization over all scales is given
by a double S\'{e}rsic model,
\begin{equation}
\begin{array}{rcl}
I_{dS}(R) & = & I_{S_1}(R) + I_{S_1}(R) \\
I_{dS}(R) & = & I_{e,1} \exp \left\{ -b_{n,1} \biggl [ \biggl ({R \over R_{e,1}} \biggl)^{1/n_1} - 1 \biggr ] \right\}
+ I_{e,2} \exp \left\{ -b_{n,2} \biggl [ \biggl ({R \over R_{e,2}} \biggl)^{1/n_2} - 1 \biggr ] \right\} \\
\end{array}
\end{equation}
in which one component 
corresponds to the galaxy profile and the other to the central light
excess (i.e., a compact spheroidal or flattened stellar component). In
this case, a total of
six parameters are needed to describe the entire brightness profile
($I_{e,1}$, $n_1$, $R_{e,1}$, $I_{e,2}$, $n_2$ and $R_{e,2}$).

Thus, for our second (``{\tt composite}") description of the ensemble profiles we adopt,
on an object-by-object basis, one of three parameterizations that best fits
the observed profile: (1) a core-S\'{e}rsic model (Eq.~2); (2) a single, unbroken
S\'{e}rsic model (Eq.~1); or (3) a double S\'{e}rsic model (Eq.~4).
As mentioned above, the precise choice of parameterization is found to
depend strongly on galaxy luminosity: all galaxies brighter
than $M_B \lesssim -20$ are modeled with core-S\'{e}rsic laws, while most
galaxies slightly fainter than this ($-20 \lesssim M_B \lesssim -19.5$) 
are well fitted with a single S\'{e}rsic model over all radii. For the majority 
of galaxies fainter than $M_B \gtrsim -19.5$, a double S\'{e}rsic model
can accurately match the profiles on all scales. A small number of
galaxies ($\lesssim$ 10\% of the sample) fainter than $M_B \approx -17.5$
are found {\it not} to require a second S\'{e}rsic component (i.e., they
contain no obvious central nucleus and their profiles are generally well
fitted by a single S\'{e}rsic model). We shall return to these interesting
objects in \S\ref{sec:discussion}. The long dashed curves in Figures~1 and 2 show
these ``composite models", with the exact choice of parametrization listed
in the lower left corner of each panel. The short and long dashed curves are
generally indistinguishable, confirming that, for these particular galaxies,
either approach yields an acceptable parameterization of the profiles on
both large and small scales. Our conclusions are therefore robust
to the choice of these two parameterizations.

\section{Results}
\label{sec:results}

\subsection{Central Surface Brightness Profile Slopes}
\label{sec:slopes}

While it is well established that the brightest early-type galaxies differ
from their fainter counterparts in terms of isophotal shape, ellipticity,
kinematics and stellar populations (see, e.g., Bender \etal 1989; Kormendy \& 
Djorgovski 1989; Caon \etal 1993; Ferrarese \etal 2006a; Emsellem \etal 2007)
the strongest evidence for a bonafide {\it dichotomy} in their structural properties
--- rather than a continous variation along the luminosity function with
some intrinsic scatter --- has come from the behavior of their central brightness
profile slopes (Ferrarese \etal 1994; Lauer \etal 1995; Rest \etal 2001; 
Ravindranath \etal 2001; Lauer \etal 2007). However, Ferrarese \etal (2006ac) noted that
such slope measurements have often relied on heterogeneous archival HST images
(i.e., different instruments, filters, and resolution limits) or
on galaxy samples with ill-defined selection functions and spanning
wide ranges in distance (e.g., from 3.5~Mpc to $\sim$ 320~Mpc in
the case of the recent study by Lauer \etal 2007). If
such a dichotomy has a physical origin, then it would seem more
appropriate to measure the slope at either the same physical --- rather
than angular --- radius, as was done in Ferrarese \etal (2006a), or at a point
corresponding to a constant fraction of some characteristic scale radius in
every galaxy (see also Ferrarese \etal 2006c).

\begin{figure}
\plotone{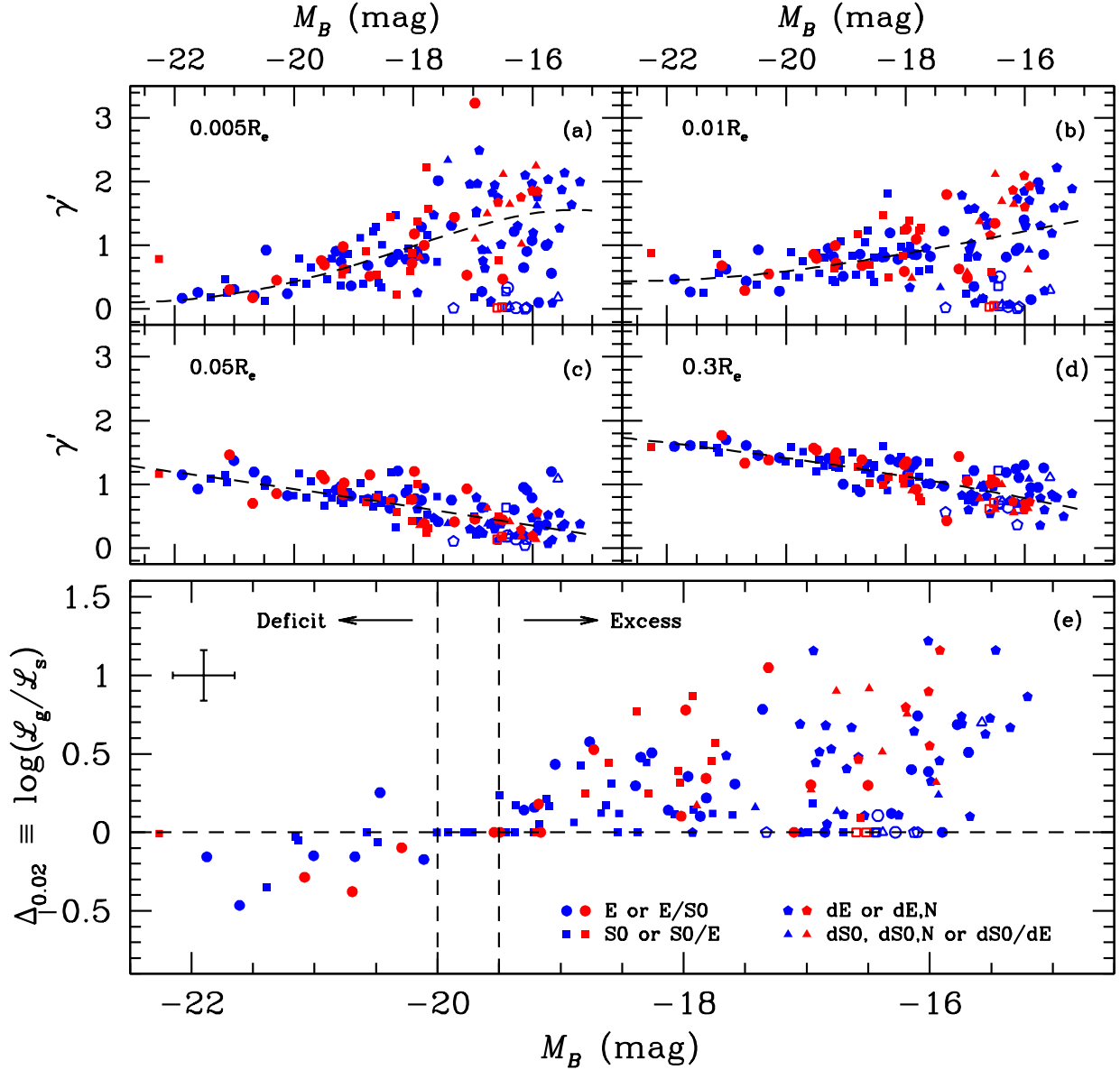}
\figurenum{3}
\caption{{\it (Panels a-d)} Slope of the best-fit composite model, $\gamma^{\prime}$, measured at
different fractions of $R_e$ (i.e., 0.005, 0.01, 0.05 and 0.30). 
Virgo and Fornax galaxies are shown by blue and red symbols, respectively. 
The behavior of the $\gamma^{\prime}$-$M_B$ relation depends sensitively on
the radius at which $\gamma^{\prime}$ is measured, but there is no evidence
for a ``core/power-law" dichotomy. The dashed curves show low-order Legendre polynomials
that highlight the trends with magnitude.
{\it (Panel e)} Dependence of $\Delta_{0.02} = \log({\cal L}_g/{\cal L}_s)$ on
galaxy magnitude for the composite fits (see text for details).
Galaxies with central luminosity deficits have $\Delta_{0.02} < 0$ while those with central
excesses have $\Delta_{0.02} > 0$. There is a smooth transition from
central deficit to excess with decreasing galaxy luminosity. A typical errorbar is
shown on the left side of the panel. Open 
symbols denote those galaxies with dE/dIrr
transition morphologies, dust, young stellar clusters and/or evidence of young stellar
populations from blue integrated colors.
\label{fig:delta}}
\end{figure}

Panels {\it (a-d)} of Figure~3 illustrate this point by plotting the slope of
the best-fit composite model at differing fractions of the effective 
radius of the galaxy (0.005, 0.01, 0.05 and 0.30) as a function of 
absolute blue magnitude.\footnote{Absolute magnitudes for the ACSVCS galaxies are computed from
the apparent~magnitudes given in C\^ot\'e \etal (2004), reddenings as 
described in Jord\'an \etal (2004) and SBF distances from Mei \etal (2007).
For the 11 Virgo galaxies lacking SBF distances, we assume $(m-M)$ = 31.09~mag.
For the ACSFCS galaxies, we use the apparent magnitudes and reddenings from Jord\'an
\etal (2007) and SBF distances from Blakeslee \etal (2008, in preparation). Four
ACSFCS galaxies without SBF distances are assigned the median Fornax distance.}
That is to say, for each galaxy in our sample, we measure the instantaneous slope
of the fitted models using the relations
\begin{equation}
\begin{array}{rcl}
\gamma^{\prime}_{\rm cS} & = & {d\log I_{cS} \over d\log R} = {\gamma\over (R/R_b)^{\alpha} + 1} + {b_n \over n}\left({R \over R_e}\right)^{\alpha} \left[{R^{\alpha}+R_b^{\alpha} \over R_e^{\alpha} }\right]^{1/(\alpha{n}) - 1} \\
\gamma^{\prime}_{\rm S} & = & {d\log I_{S} \over d\log R} = - {b_n \over n}\left(R \over R_e \right)^{1/n} \\
\gamma^{\prime}_{\rm dS} & = &  {d\log I_{dS} \over d\log R} = {1 \over I_{S_1}(R) + I_{S_2}(R)}\bigg[I_{S_1}(R){b_{n,1} \over {n_1}}\left(R \over R_{e,1} \right)^{1/n_1} + I_{S_2}(R){b_{n,2} \over n_2}\left(R \over R_{e,2} \right)^{1/n_2} \bigg] \\
\end{array}
\end{equation}
where the {\tt cS}, {\tt S} and {\tt dS} subscripts refer to the core-S\'{e}rsic, S\'{e}rsic
and double-S\'{e}rsic parameterizations discussed in \S\ref{sec:param}. Note that this
approach differs from the one adopted by Ferrarese \etal (2006a) who --- recognizing that 
compact stellar mass concentrations become increasingly prominent among progressively fainter
and lower surface brightness galaxies --- preferred to measure the slope of the underlying 
{\it galaxy component} rather
than using the combined (i.e., galaxy + nucleus) profiles. The photometric and structural scaling
relations derived by Ferrarese \etal (2006a) corroborated previous reports of a continuum in 
galaxy properties fainter than $M_B \approx -20$ (e.g., see \S4 and 5 of
Ferrarese \etal 2006a and references therein), with no dichotomy between giants
and dwarfs (see also the discussion in \S1 of Graham \& Guzman 2003).

Panels {\it (a-d)} show the $\gamma^{\prime}$-$M_B$ relations found when the 
profiles are parameterized as described in \S 2: a core-S\'ersic profile (Eq.~2) for
the brightest galaxies, a double S\'ersic profile (Eq.~4) for most of the low- and
intermediate-luminosity galaxies, and a S\'ersic profile in all other cases (mostly
intermediate luminosity and a few low-luminosity galaxies).  
Four important points are illustrated in these panels. First, the
overall trends defined by the Virgo and Fornax galaxies are indistinguishable. Second,
the behavior of the $\gamma^{\prime}$-$M_B$ relation\footnote{We present these scaling relations in terms of $M_B$ for two
reasons. First, the scaling relations and dichotomies discussed in this paper have
traditionally been expressed in this bandpass. Second, placing the ACSVCS and
ACSFCS galaxies in the broader context of their cluster environments
(i.e., see \S\ref{sec:lauer}) is most straighforward using this bandpass since
the wide-field,
photographic (blue emulsion) surveys of Binggeli, Sandage \& Tammann (1985) and Ferguson (1989) remain
the most homogeneous photometric catalogs for Virgo and Fornax cluster galaxies.
Homogeneous `curve-of-growth' photometry from wide-field {\it ugrizJHK} imaging
for the ACSVCS galaxies will be presented in Chen \etal (2008, in preparation).} is obviously sensitive to the exact
choice of measurement radius, particularly on the smallest
angular scales.
Third, the $\gamma^{\prime}$-$M_B$ relation undergoes an unmistakable ``inversion"
between $\sim 0.01$-$0.05R_e$: on the smallest scales,
the profiles are found to {\it steepen
systematically} as one moves down the luminosity function, whereas 
on larger scales, the profiles become {\it progressively
shallower} with decreasing luminosity.
And, finally, in no case is there clear evidence for a dichotomy or bimodality
(see also Figures~1 and 2).

\subsection{Comparison to Lauer et al. (2007)}
\label{sec:lauer}

The continuity of the $\gamma^{\prime}$-$M_B$ relations shown in the first four panels of Figure~\ref{fig:delta},
while consistent with the finding of no core/power-law dichotomy by Ferrarese \etal (2006a), 
is apparently at odds with a recent detection of strong bimodality in $\gamma^{\prime}$
by Lauer \etal (2007). However, it is clear from this figure that the distribution of 
slopes found for any sample of galaxies will depend sensitively on both the
choice of measurement radii and the luminosity distribution of the galaxies themselves.

\begin{figure}
\plotone{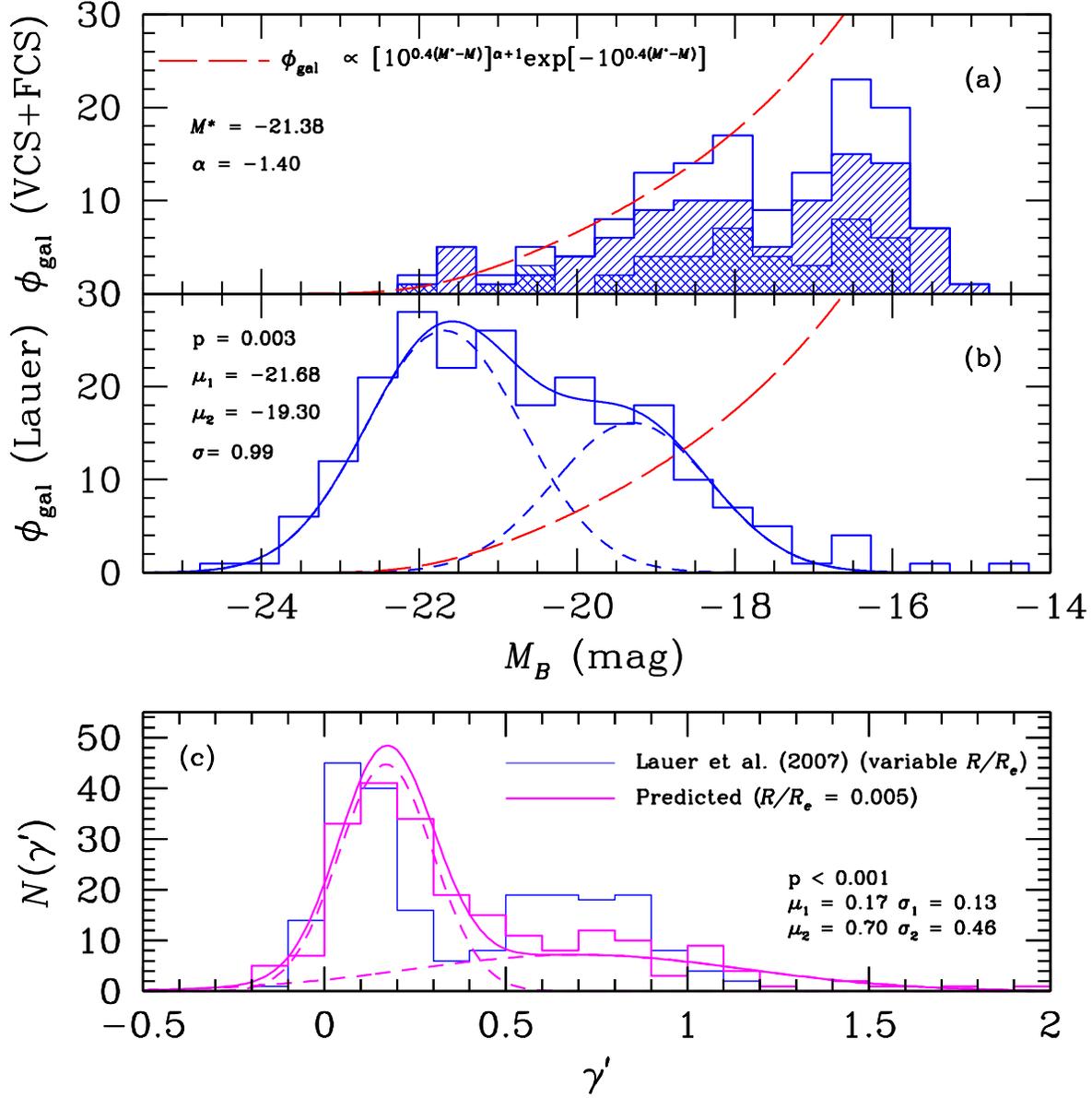}
\figurenum{4}
\caption{{\scriptsize{
{\it (Panel a)} Luminosity function, $\phi_{\rm gal}$, for the combined sample of 143 early-type galaxies 
from the ACS Virgo and Fornax Cluster Surveys (histogram). The dashed red curve shows a Schechter function with
parameters $\alpha = -1.40$ and  $B^* = 9.8$ ($M_B^* \approx -21.4$), appropriate for E+S0+dE+dS0
galaxies in the Virgo Cluster (Sandage, Binggeli \& Tammann 1985). The normalization is chosen to match the program galaxy luminosity
function at $M_B \lesssim -19$ since both the Virgo and Fornax samples are complete above this luminosity.
{\it (Panel b)} Luminosity function for the 219 galaxies in the compilation of Lauer \etal (2007). The sample is
strongly non-representative
of a Schechter function (the dashed curve from {\it panel (a)} is reproduced for comparison). A KMM
analysis shows the sample to be bimodal in luminosity, with peaks at $M_B = -21.68$ and $-19.30$, and
a 0.3\% chance of drawing these data randomly from a unimodal Gaussian distribution (the null hypothesis). The
dotted and solid curves show the separate Gaussians, and their sum, as determined by KMM.
{\it (Panel c)} A comparison of the distribution of $\gamma^{\prime}$ values measured by Lauer \etal
(2007) (thin blue histogram) with that expected (magnenta histogram) based on their program galaxy
luminosities and the $\gamma^{\prime}$-$M_B$ relation measured at $R = 0.005R_e$ (see Figure~3, panel a). 
KMM reveals there to be a less than 0.1\% chance of drawing these data randomly from a unimodal Gaussian
distribution.}}
\label{fig:bimodal}}
\end{figure}

In the upper panel of Figure~\ref{fig:bimodal}, we plot as the open blue histogram
the luminosity function, $\phi_{\rm gal}$, of the
143 galaxies that make up the ACS Virgo and Fornax Cluster Surveys. The single- and
double-hatched histograms show the results for the ACSVCS and ACSFCS samples, respectively.
The dashed red curve shows a Schechter function with $\alpha = -1.40$ and $B^* = 9.8$
($M_B^* = -21.4$), the best-fit parameters for early-type (E+S0+dE+dS0) galaxies
in the Virgo Cluster according to Sandage, Binggeli \& Tammann (1985). The normalization of the
Schechter function has been chosen to match the luminosity function of the ACSVCS
and ACSFCS sample
galaxies brighter than $M_B \approx -19$. Recall that both surveys are complete
above this level and that the ACSFCS sample is complete for $M_B \lesssim -16.1$.

The blue histogram in {\it panel (b)} shows the luminosity distribution of the 219 galaxies analysed by
Lauer \etal (2007).\footnote{For comparison to the ACSVCS and ACSFCS samples, we have converted the $M_V$ magnitudes
given in Lauer \etal (2007) to $M_B$ by assuming $(B-V) = 0.96$ for BCG/E galaxies
and $(B-V) = 0.85$ for S0 galaxies (Fukugita, Shimasaku \& Ichikawa 1995). A color
of $(B-V) = 0.85$ was also adopted for the lone Sa galaxy in their sample.}
The Schechter function
from {\it panel (a)} is reproduced as the dashed red curve.
The common element in the selection of these galaxies was the
availability in the literature of Nuker model fits to brightness profiles derived 
from either WFPC1, WFPC2, NIC2 or NIC3 imaging. Two important properties of the
Lauer \etal (2007) sample are worth noting. First, it is unrepresentative of
the Schechter function form that provides a reasonable match to the luminosity distribution of
the early-type galaxy populations in Virgo and Fornax in particular (Sandage \etal 1985; 
Ferguson \& Sandage 1988) 
and, more generally, those of galaxies in both cluster or field enviroments
(e.g., Schechter 1976; Loveday \etal 1992; Marzke \etal 1994; Blanton \etal 2003). Second,
as Lauer \etal (2007) point out, their sample is bimodal in luminosity --- a result that we
confirm. A KMM analysis (McLachlan \& Basford 1988; Ashman, Bird \&
Zepf 1994) indicates a 0.3\% chance that these data are drawn randomly from a
unimodal Gaussian distribution (the null hypothesis). Assuming homoscedasticity
and clipping seven galaxies with $M_B > -17.5$ (which serve to make the distribution
even more bimodal), we find peaks at $M_B = -21.68$ and $-19.30$. 

What then is the expected distribution of $\gamma^{\prime}$ values for this dataset? As
{\it panels (a-d)}
of Figure~\ref{fig:delta} demonstrate, the detailed form of the $\gamma^{\prime}$ distribution 
will depend not only on the program galaxy magnitudes, but also on the choice
of radius, $R/R_e$, at which $\gamma^{\prime}$ is measured. Lauer \etal (2007) choose to measure $\gamma$ at the
data resolution limit (0\farcs02, 0\farcs04 or 0\farcs1 depending on the instrument) for
all galaxies in their sample, regardless of their distance or effective radius. 
According to the data in their Tables~1 and 2, $R/R_e$ thus varies 
between 0.00017 and 0.030 (a factor of 178) for the 150 galaxies in their sample
for which estimates of $R_e$ are available in their Table~1. To illustrate the
behavior of the $\gamma^{\prime}$ distribution expected for their sample, we
combine their $M_B$ values, calculated as described above, with the $\gamma^{\prime}$-$M_B$
relation shown as the smooth polynomial in {\it panel (a)} of Figure~\ref{fig:delta}: i.e.,
for $R/R_e = 0.005$, which falls within the range of $R/R_e$ values used by Lauer \etal (2007).
To approximate intrinsic scatter and measurement uncertainties, we assume a fixed
dispersion of $\sigma$($\gamma^{\prime}$) = 0.1 for $M_B < -22$, and rising
linearly with galaxy magnitude to $\sigma$($\gamma^{\prime}$) = 0.5 at $M_B=-16$.

The $\gamma^{\prime}$ distribution found in a typical simulation is
 shown by the heavy magenta histogram
in {\it panel (c)}. In this case, KMM identifies highly significant peaks at
$\gamma^{\prime} \approx 0.17$ and $0.70$,
with a less than 0.1\% chance of drawing 
these data from a unimodal Gaussian distribution. Thus, the simulated distribution
is found to be strongly bimodal {\it despite the fact that the input
$\gamma^{\prime}$-$M_B$ varies smoothly along the luminosity function}.
The peak values of the simulated distribution should be compared
to the values of 0.11 and 0.70 that we find when applying KMM to the
actual $\gamma^{\prime}$ measurements tabulated in Table~2 of Lauer \etal (2007),
reproduced here as the thin blue histogram. Although the two distributions differ in the details 
(which is not surprising since, as mentioned above, Lauer \etal do not measure $\gamma^{\prime}$
at a fixed fraction of $R_e$ or any other fixed physical scale), our exercise suggests that
the bimodality in $\gamma^{\prime}$ reported by Lauer \etal (2007) is
likely an artifact of a sample defined by
a bimodal luminosity distribution that is: (1) unrepresentative of real galaxy ensembles;
and (2) strongly biased in favor of bright, but intrinsically rare, galaxies that are known
to have nearly constant-density cores. For instance, 56 of the galaxies in the Lauer et al.
compilation (26\% of their sample) are brighter than VCC1226 (M49 = NGC4472), the brightest
member of the Virgo Cluster.

\subsection{A Characteristic Radius: $R_b$}
\label{sec:radii}

To first order, S\'{e}rsic models provide accurate
representations of the global, azimuthally averaged brightness profiles for 
almost all galaxies in our sample, 
largely independent of luminosity, morphological type, prior classification
as giant or dwarf, and the presence or absence of morphological pecularities
such as rings, shells and bars. However, the central deviations provide a notable
exception to this rule. It is striking that, in all galaxies, the inner
departure from the S\'{e}rsic model occurs at very nearly the same fraction
of the effective radius. Using the core-S\'{e}rsic
parameterization (which can be applied uniformly to all galaxies)
we find the median logarithmic ratio of the break radius to the effective
radius to be $\langle\log(R_b/R_e)\rangle = -1.68$ with $rms$ scatter 0.33. 
Thus, the inner deviations from the S\'{e}rsic law are found to occur
at a characteristic radius of $R_b \sim 0.02R_e$, with some 
scatter (the $\pm$ 1$\sigma$ range is 0.010--0.045) that does not
appear to correlate with any galaxy property. As an illustration,
the vertical dashed lines in each panel of Figures~1 and 2 are drawn at 
0.02$R_e$, where $R_e$ is measured independently in the $g$ and $z$
profiles. 

\subsection{Light Deficits and Excesses: The ${\Delta}_{0.02}$ Parameter}
\label{sec:delta}

Because the departures from the inward extrapolation of the outer S\'{e}rsic
component occur at $R_b \approx 0.02R_e$, we define a parameter, ${\Delta}_{0.02}$,
as the logarithm of the ratio of the total luminosity of the
best-fit galaxy model (core-S\'ersic or composite), ${\cal L}_g$, inside this
radius to that of the outer S\'{e}rsic component in this same region:
${\Delta}_{0.02} = \log{({\cal L}_g/{\cal L}_s})$. Note 
that this description of the central behavior of the brightness profiles
has the advantage relative to $\gamma^{\prime}$ that it is an {\it integral}
measurement, rather than a differential one, and is thus less 
susceptible to noise in the profiles, PSF uncertainties, and 
the extent to which the stellar nucleus may be resolved. 
Galaxies with central light
deficits then have ${\Delta}_{0.02} < 0$, those with excesses have
${\Delta}_{0.02} > 0$, and those whose profiles are
well represented by S\'{e}rsic models over the full range
in radius have ${\Delta}_{0.02} = 0$.

\begin{figure}
\plotone{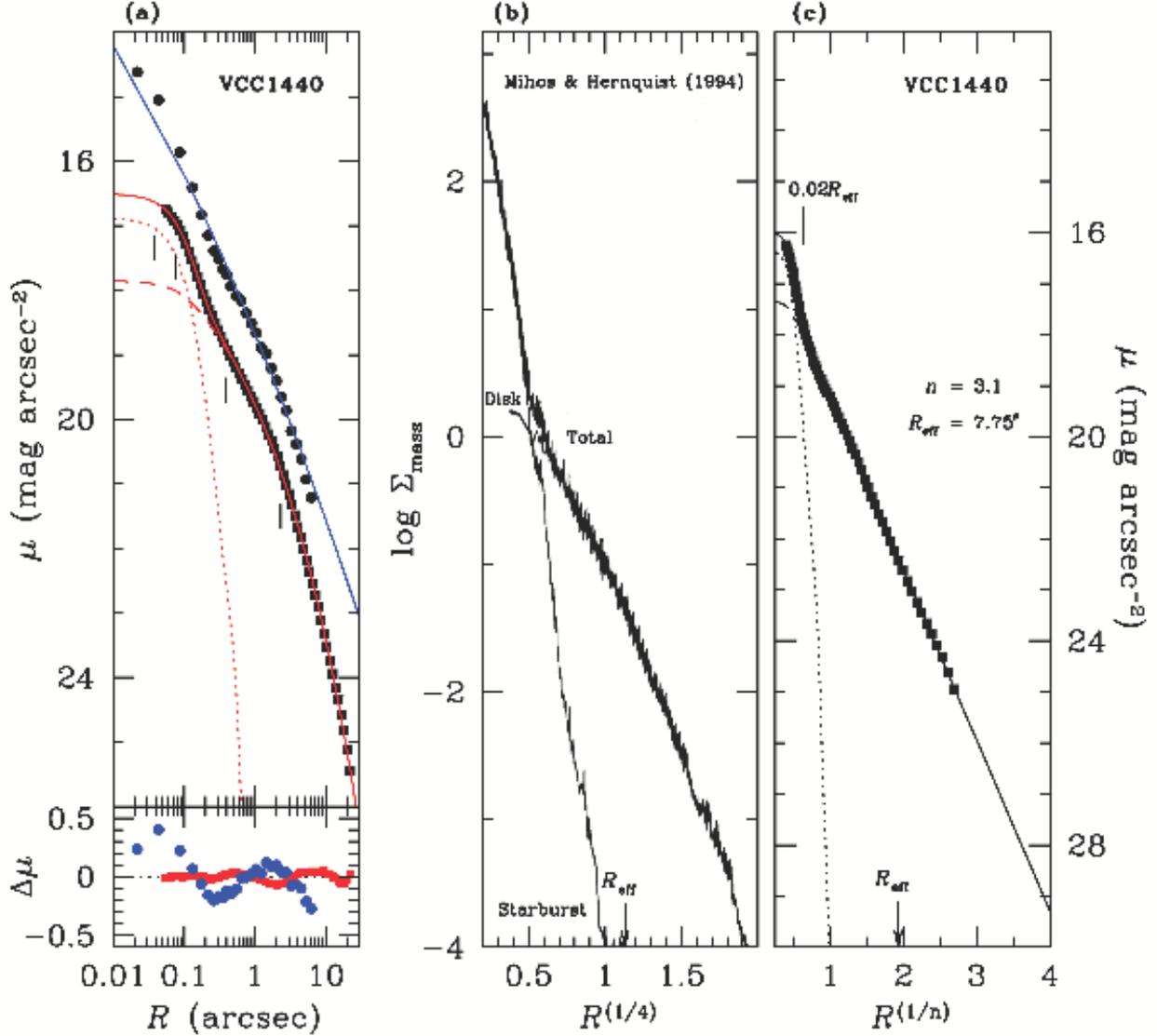}
\figurenum{5}
\caption{{\scriptsize{
{\it (Panel a)} Comparison of WFPC1 and ACS/WFC brightness profiles for VCC1440,
an E0 galaxy ($M_B = -15.94$) previously classified as having a ``power-law" profile
(Lauer \etal 1995; 2007).
The upper datapoints show the deconvolved, major axis, F555W WFPC1
profile of Lauer et~al. (1995) and their fitted Nuker model. The lower datapoints
show the azimuthally averaged ACS F475W profile along with the PSF-convolved,
double S\'{e}rsic model (dashed and dotted curves) whose sum (solid red curve) best
fits the observed profile. The datasets and models have been shifted apart
by 0.75~mag for clarity. Residuals for
both parameterizations are shown in the lower panel; note the {\tt S}-shaped
residuals for the Nuker model, indicative of a central nucleus (blue symbols; see also Ferrarese \etal 2006c).
{\it (Panel b)} Predicted luminous mass profiles for the remnants of merging disk/halo
galaxy models, showing the ``dense stellar core" that forms as a result of gas
dissipation and star formation. {\it Panel (b)} is adapted from the data/simulations shown in
Figure~1a of Mihos \& Hernquist (1994).
{\it (Panel c)} Alternate presentation of the ACS brightness profile for VCC1440,
showing the excess light inside $\approx$ 0.02$R_e$
(${\Delta}_{0.02} = 0.387\pm0.006$). Compare with {\it panel (b)}.
In {\it panels (b)} and {\it (c)}, the abscissa spans the same range when
normalized to $R_e$ ($10^{-3} \lesssim R/R_e \lesssim 10$).}}
\label{fig:v1440}}
\end{figure}

Panel {\it (e)} of Figure~3 shows the dependence of ${\Delta}_{0.02}$ on absolute
blue magnitude for the sample galaxies. The overall
trends exhibited by the Virgo and Fornax galaxies are the same: a
gradual transition from light deficits ($-0.5 \lesssim {\Delta}_{0.02} \lesssim 0$)
for galaxies brighter than $M_B \approx -20$ to light excesses
($0 \lesssim {\Delta}_{0.02} \lesssim 1.2$) for the faintest program
galaxies. This description of the inner galaxy
also allows a straightforward definition of nucleated vs. non-nucleated
galaxies. For instance, adopting ${\Delta}_{0.02} > 0$ as the
definition of a central stellar nucleus, one finds
69\% of the combined sample galaxies, and
82\% of all galaxies fainter than $M_B \approx -19$, 
to be nucleated. These estimates are consistent with the values
for ACSVCS galaxies reported in C\^ot\'e \etal (2006).

We point out that the tendency for early-type galaxies of low- and intermediate luminosity to
have central brightness profiles that fall above S\'{e}rsic laws is not a new result: several
previous investigators had noted that such behavior in isolated galaxies or using small datasets (e.g., Binggeli \&
Jerjen 1998; Kormendy 1999; Stiavelli \etal 2001; Graham \& Guzm\'an 2003). However, the 
systematic variation along the luminosity function from central light deficit to light excess ---
with no apparent structural dichotomies between ``core" and ``power-law" galaxies at $M_B \approx -20$,
nor between giants
and dwarfs at $M_B \approx -18$ (c.f. Kormendy 1985) --- has only been possible thanks to the
uniformity, homogeneity and luminosity coverage of the ACSVCS and ACSFCS datasets. 

\section{Discussion}
\label{sec:discussion}

The central deficits in bright ellipticals have been traditionally
interpreted as evidence for core evacuation by coalescing supermassive black
hole (SBH) binaries (Ebisuzaki \etal 1991; Faber \etal 1997; Milosavljevi\'c \& Merritt 2001),
although at least one recent study suggests that the dissipationless collapse of initially cold stellar distributions in existing dark matter haloes can also give rise to such features (Nipoti et al. 2006).
The discovery 
of a systematic transition from central deficit to excess should provide an
important new clue to the formation of both high- and low-luminosity early-type
galaxies and the connection to SBHs. It is worth noting that the 
mass fraction contributed by the central nuclei in low- and intermediate-luminosity
galaxies is indistinguishable ($\sim 0.2$\%) from that of SBHs in bright galaxies,
suggestive of a shared formation path (C\^ot\'e \etal 2006, Ferrarese \etal 2006ab;
Wehner \& Harris 2006; Rossa \etal 2006). On the other hand, McLaughlin \& King (2006)
point out that feedback by stellar winds and supernovae may also be capable of 
reproducing the observed scaling relations without a direct connection between the
nuclei and SBHs. And, as Ferrarese \etal (2006b) note,  there are at least some galaxies
(such as M32) that contain both a central excess (Worthey 2004) and a SBH (Verolme \etal 2002).

A lengthy discussion of the selection effects that prevented the ubiquitous
nature of these nuclei from being fully appreciated with ground-based imaging was
presented in C\^ot\'e \etal (2006). Not surprisingly, the foremost advangtage of 
HST/ACS in this regard is its exceptional angular resolution. One might think, then,
that the detection of
nuclei in low- and intermediate-luminosity would have been straightforward in early
high-resolution HST imaging, but this often proved not to be the case. For instance, the
left panel of Figure~\ref{fig:v1440} compares our ACS profile for VCC1440 --- a
low-luminosity elliptical previously classified as a ``power-law" galaxy ---
to the WFPC1 profile of Lauer \etal (1995, 2007).\footnote{Lauer \etal (2007) 
report $\gamma^{\prime} = 0.89$ for this galaxy. We measure
$\langle\gamma^{\prime}\rangle = 1.08, 1.40, 0.79$ and $1.31$ at $R/R_e = 0.005, 0.01, 0.05$
and $0.3$, respectively. These radii are indicated by the tickmarks in
Figure~\ref{fig:v1440}.} In this case, the limited
radial extent of the WFPC1 profile, and the adopted parameterization
of the galaxy profile as a power-law, allowed the full extent of the central
excess to go unrecognized, although its signature is evident in the 
{\tt S}-shaped residuals of the fitted Nuker model (see also Ferrarese \etal 2006c).

In hindsight, the realization that compact stellar mass concentrations reside at the
centers of most low- and intermediate galaxies should perhaps have come as no 
surprise. Many theoretical studies have shown that such mass
concentrations may be a generic outcome of the galaxy formation process.
Proposed formation routes include 
repeated mergers of star clusters drawn to the galaxy center via
dynamical friction, the growth of $\rho \propto r^{-7/4}$ density
cusps due to two-body relaxation of stars orbiting a central SBH
(Bahcall \& Wolf 1976)
and/or central star formation driven by gas inflows --- likely modulated by
stellar evolution, mergers, IGM confinement, and feedback from stellar
winds, supernovae or pre-existing AGN (e.g., Barnes \& Hernquist 1991; Mihos \&
Hernquist 1994; McLaughlin \etal 2006; Li \etal 2007; Hopkins \etal 2007).

A comparison of the properties (e.g., sizes, luminosity functions, mass fractions, colors)
of nuclei and star clusters in ACSVCS program 
galaxies argues against the first scenario as the {\it dominant} formation mechanism
(e.g., see \S5.2 of C\^ot\'e \etal 2006). Likewise, Bahcall-Wolf ``cusps" generated
by SBHs do not seem viable since most of the nuclei are {\it resolved} and thus much
more extended than the predicted cusps (e.g., Merritt \& Szell 2006).
On the other hand, gas inflow models --- either internally or externally modulated
--- seem more attractive as they have long predicted the formation of ``dense stellar
cores" in early-type galaxies. Consider the center and right
panels of Figure~\ref{fig:v1440} which compare the brightness profile of VCC1440 (the
same galaxy shown in the left panel)
to the simulations of Mihos \& Hernquist (1994). These authors
showed that compact central starbursts are found in the remnants of disk/halo
mergers and went on to argue that this central component would lead to ``a break
in the mass profile at small radii ($\sim 2\%R_e$)".
As Figure~\ref{fig:v1440} and
the discussion in \S\ref{sec:radii} show, this may turn out to be a remarkably
prescient prediction. The fact that the deviations from the fitted S\'ersic models
occur on roughly the same ($\sim 2\%R_e$) scale for both the central excesses {\it and}
central deficits (which are usually attributed to binary SBH evolution; see above)
presents something of a puzzle since it is unclear why these two unrelated physical 
mechanisms would give rise to deviations on the same (fractional) scale. 

In any case, it will be important to test the gas inflow scenario envisaged
by Mihos \& Hernquist (1994)
by measuring star formation and chemical enrichment histories for the
central regions of bright (deficit) and faint (excess) galaxies alike. Both
the measured color profiles (Ferrarese \etal 2006a) and the model-dependent integrated
colors (C\^ot\'e \etal 2006) of the nuclei and galaxies
in the ACSVCS suggests that stellar population differences between the two
components may indeed exist, with the nuclei in the faintest systems generally
being bluer than the underlying galaxy. On the other hand, at least some 
intermediate-luminosity galaxies show evidence for surprisingly
red central components (e.g., $g-z \gtrsim 1.5$). 

Finally, we point out that the open symbols in {\it panel (e)} of Figure~\ref{fig:delta}
denote galaxies with
``dE/dIrr transition" morphologies: i.e., dust, young stellar clusters and/or evidence
of young stellar populations from blue integrated colors (Ferrarese \etal 2006a;
C\^ot\'e \etal 2006). Such galaxies
typically lack conspicuous central nuclei at their photocenters. The absence of a central excess
may suggest that the process of nucleus-building --- perhaps
through delayed gas inflows and subsequent star formation --- has yet to occur in
these apparently young galaxies.

\acknowledgments

Support for programs GO-9401 and GO-10217 was provided through grants from
STScI, which is operated by AURA, Inc., under NASA contract NAS5-26555. 
Additional support for P.C. was provided by NASA LTSA grant NAG5-11714.
C.W.C. acknowledges support provided by National Science Council of Taiwan.
L.I. acknowledges support from Fondap Center of Astrophysics.
M.J.W. acknowledges support through NSF grant AST-0205960.
This research has made use of the NASA/IPAC Extragalactic Database (NED)
which is operated by the Jet Propulsion Laboratory, California Institute
of Technology, under contract with the National Aeronautics and Space Administration.
This publication also makes use of data products from the Sloan Digital Sky Survey (SDSS).
Funding for SDSS and SDSS-II has been provided by the Alfred P. Sloan Foundation, the
Participating Institutions, the National Science Foundation, the U.S. Department of Energy,
the National Aeronautics and Space Administration, the Japanese Monbukagakusho, the
Max Planck Society, and the Higher Education Funding Council for England.

Facilities: HST(ACS/WFC)


\begin{thebibliography}{}
\bibitem[]{} Adelman-McCarthy, J., \etal 2007, \apjs, in press, VizieR Online Data Catalog, 2276, 0
\bibitem[]{} Ashman, K.M., Bird, C.M., \& Zepf, S.E.\ 1994, \aj, 108, 2348 
\bibitem[]{} Bahcall, J.N., \& Wolf, R.A. 1976, \apj, 209, 214 
\bibitem[]{} Barnes, J.E., \& Hernquist, L. 1991, \apj, 370, L65
\bibitem[]{} Bender, R. \etal 1989, \aap, 217, 35
\bibitem[]{} Binggeli, B., Sandage, A., \& Tammann, G.A. 1985, \aj, 90, 1681
\bibitem[]{} Binggeli, B., \& Jerjen, H.\ 1998, \aap, 333, 17
\bibitem[]{} Blanton, M.R., et al.\ 2003, \apj, 592, 819 
\bibitem[]{} Caon, N., Capaccioli, M. \& D'Onofrio, M. 1993, \mnras, 265, 1013
\bibitem[]{} C\^ot\'e, P., \etal\  2004, \apjs, 153, 223 (ACSVCS Paper~I)
\bibitem[]{} C\^ot\'e, P., \etal\  2006, \apjs, 165, 57 (ACSVCS Paper~VIII)
\bibitem[]{} Ebisuzaki, T., Makino, J., \& Okumura, S.K.\ 1991, \nat, 354, 212
\bibitem[]{} Emsellem, E., et~al.\ 2007, \mnras, in press, ArXiv Astrophysics e-prints, arXiv:astro-ph/0703531 
\bibitem[]{} Faber, S.M., et~al.  1997, \aj, 114, 1771 
\bibitem[]{} Ferguson, H.C.\ 1989, \aj, 98, 367 
\bibitem[]{} Ferguson, H.C. \& Sandage, A. 1988, \aj, 96, 1520
\bibitem[]{} Ferrarese, L., van den Bosch, F.C., Ford, H.C., Jaffe, W., \& O'Connell, R.W. 1994, \aj, 108, 1598
\bibitem[]{} Ferrarese, L., \etal\ 2006a, \apjs, 164, 334 (ACSVCS Paper~VI)
\bibitem[]{} Ferrarese, L., \etal\ 2006b, \apj, 644, L21
\bibitem[]{} Ferrarese, L., \etal\ 2006c, ArXiv Astrophysics e-prints, arXiv:astro-ph/0612139 
\bibitem[]{} Ford, H.C., \etal\ 1998, Proc. SPIE, 3356, 234
\bibitem[]{} Fukugita, M., Shimasaku, K., \& Ichikawa, T.\ 1995, \pasp, 107, 945 
\bibitem[]{} Graham, A.W. 2004, \aj, 613, L33
\bibitem[]{} Graham, A.W., Erwin, P., Trujillo, I., \& Asensio Ramos, A. 2003, \aj, 125, 2951
\bibitem[]{} Graham, A.~W., \& Guzm{\'a}n, R.\ 2003, \aj, 125, 2936 
\bibitem[]{} Grant, N.I., Kuipers, J.A., \& Phillipps, S. 2005, \mnras, 363, 1019
\bibitem[]{} Hopkins, P.~F., Hernquist, L., Cox, T.~J., Robertson, B., \& Krause, E.\ 2007, ApJ, submitted,
ArXiv Astrophysics e-prints, arXiv:astro-ph/0701351 
\bibitem[]{} Jord\'an, A., et~al.\ 2004, \apjs, 154, 509 (ACSVCS Paper~II)
\bibitem[]{} Jord\'an, A., et~al.\ 2007, \apjs, 169, 213 (ACSFCS Paper~I)
\bibitem[]{} Kormendy, J. 1985, \apj, 295, 73
\bibitem[]{} Kormendy, J., \& Djorgovski, S. 1989, \araa, 27, 235
\bibitem[]{} Kormendy, J.\ 1999, Galaxy  Dynamics - A Rutgers Symposium, 182, 124 
\bibitem[]{} Lauer, T.R., et~al. 1995, \aj, 110, 2622
\bibitem[]{} Lauer, T.R., et~al. 2007, \apj, 664, 226
\bibitem[]{} Li, Y., Haiman, Z., \& Mac Low, M.-M.\ 2007, ApJ, 663, L61
\bibitem[]{} Loveday, J., Peterson, B.A., Efstathiou, G., \& Maddox, S.J.\ 1992, \apj, 390, 338 
\bibitem[]{} Marzke, R.O., Huchra, J.P., \& Geller, M.J.\ 1994, \apj, 428, 43 
\bibitem[]{} McLachlan, G.J., \& Basford, K.E.\ 1988, Mixture Models: Inference and Application to Clustering (New York: Dekker)
\bibitem[]{} McLaughlin, D.E., King, A.R., \& Nayakshin, S. 2006, \apj, 650, L37
\bibitem[]{} Mihos, J.C., \& Hernquist, L.\ 1994, \apj, 437, L47
\bibitem[]{} Milosavljevi\'c, M., \& Merritt, D. 2001, \apj, 563, 34
\bibitem[]{} Mei, S., et~al.\ 2005, \apj, 625, 121 (ACSVCS Paper~V)
\bibitem[]{} Mei, S., et~al.\ 2007, \apj, 655, 144 (ACSVCS Paper~XIII)
\bibitem[]{} Merritt, D., \& Szell, A. 2006, \apj, 648, 890 
\bibitem[]{} Nipoti, C., Londrillo, P., \& Ciotti, L. 2006, \mnras, 370, 681
\bibitem[]{} Ravindranath, S., Ho, L.C., Peng, C.Y., Filippenko, A.V., \& Sargent, W.L.W. 2001, \aj, 122, 653
\bibitem[]{} Rest, A., \etal 2001, \aj, 121, 2431
\bibitem[]{} Rossa, J., \etal 2006, \aj, 132, 1074
\bibitem[]{} Sandage, A., Binggeli, B., \& Tammann, G.A. 1985, \aj, 90, 1759 
\bibitem[]{} S\'ersic, J.-L. 1968, Atlas de Galaxias Australes (C\'ordoba: Obs. Astron., Univ. Nac. C\'ordova) 
\bibitem[]{} Schechter, P.\ 1976, \apj, 203, 297 
\bibitem[]{} Schweizer, F.\ 1980, \apj, 237, 303 
\bibitem[]{} Stiavelli, M., Miller, B.W., Ferguson, H.C., Mack, J., Whitmore, B.C., \& Lotz, J.M.\ 
2001, \aj, 121, 1385
\bibitem[]{} West, A.A., \etal 2007, \aj, submitted
\bibitem[]{} Verolme, E.K., et al.\ 2002, \mnras, 335, 517 
\bibitem[]{} Wehner, E.H., \& Harris, W.E. 2006, \apj, 644, L17
\bibitem[]{} Worthey, G.\ 2004, \aj, 128, 2826 
\end{thebibliography}
\end{document}